\documentclass[aps,pra,floatfix,superscriptaddress,showpacs,twocolumn]{revtex4}
\usepackage{amsmath}

\usepackage{graphicx}
\usepackage{dcolumn}
\usepackage{bm}
\usepackage{euscript}

\input psfig.sty

\def\prn#1{{\left(#1\right)}}

\def\sbrk#1{{\left[#1\right]}}

\def\bra#1{{\langle#1|}}

\def\cg(#1,#2)(#3,#4)(#5,#6){\bra{#1,#2,#3,#4}#5,#6\rangle}
\def\threej(#1,#2)(#3,#4)(#5,#6){\begin{pmatrix}#1&#3&#5\\#2&#4&#6\end{pmatrix}}
\def\sixj(#1,#2,#3)(#4,#5,#6){\begin{Bmatrix}#1&#2&#3\\#4&#5&#6\end{Bmatrix}}
\def\ninej(#1,#2,#3)(#4,#5,#6)(#7,#8,#9){\begin{Bmatrix}#1&#2&#3\\#4&#5&#6\\#7&#8&#9\end{Bmatrix}}

\def\sE{{\ensuremath{\EuScript E}}}

\def\sN{{\ensuremath{\EuScript N}}}

\begin{document}

\pagenumbering{arabic}

\setcounter{footnote}{0}

\title{On the sensitivity of condensed-matter P- and T-violation experiments}

\author{D. Budker}
 \email{budker@berkeley.edu}
 \affiliation{Department of Physics, University of California at
Berkeley, Berkeley, California 94720-7300}
 \affiliation{Nuclear Science Division, Lawrence
 Berkeley National Laboratory, Berkeley, California 94720}

\author{S. K. Lamoreaux}
\email{lamore@lanl.gov}
\affiliation{University of California, Los Alamos National Laboratory, \\
   Physics Division, P-23, MS-H803, Los Alamos, New Mexico 87545}

\author{A. O. Sushkov}
 \email{alex000@socrates.berkeley.edu}
 \affiliation{Department of Physics, University of California at
Berkeley, Berkeley, California 94720-7300}

\author{O. P. Sushkov}
\email{sushkov@phys.unsw.edu.au}
  \affiliation{School of Physics, University of New South Wales,
Sydney 2052, Australia}

\date{\today}
\begin{abstract}
Experiments searching for parity- and
time-reversal-invariance-violating effects that rely on measuring
magnetization of a condensed-matter sample induced by application of
an electric field are considered. A limit on statistical sensitivity
arises due to random fluctuations of the spins in the sample. The
scaling of this limit with the number of spins and their relaxation
time is derived. Application to an experiment searching for nuclear
Schiff moment in a ferroelectric is discussed.
\end{abstract}


\maketitle

\section{Introduction}
\label{Section_Intro}

Much of the present knowledge about the fundamental symmetries CP
(invariance under combined operations of spatial inversion and
charge reversal) and T (invariance with respect to time reversal)
comes from experiments measuring P- and T-violating permanent
electric-dipole moments (EDM) of atoms, molecules, and the neutron,
see, for example, Ref. \cite{Khr97}. Most EDM experiments measure
precession of the angular momentum of the system in an applied
electric field analogous to the Larmor precession in an applied
magnetic field.

In addition to such \emph{precession experiments}, there are EDM
searches of another kind \cite{Sha68,Vas78},
which have drawn recent renewed attention
\cite{Lam2002,Liu2004,Muk2005,Der2005}.
The idea of these experiments is the following.

Suppose that we have some condensed-matter sample that has $\sN$
spins (either electron or nuclear depending on the specific
experiment). If an electric field is applied to the sample, it
interacts with the associated (P- and T-violating) EDM leading to a
slight orientation of the spins in the direction of the electric
field. This orientation, in turn, is measured by measuring the
induced magnetization of the sample.

\section{Comparison of the precession EDM experiments with
condensed-matter experiments in the high-temperature limit}
\label{Section_High-Temp_Limit}

The signal in ``traditional" EDM experiments is given by
\begin{align}
S_1 \approx \sN \frac{dE}{\hbar}\,\tau.\label{Eq_EDM_Sens_trad_S1}
\end{align}
Here $\sN$ is the number of particles involved in a measurement, $d$
is the EDM, $E$ is the effective electric field acting on the
particle, and $\tau$ is the spin-relaxation time. This assumes a
``single-shot Ramsey-type" measurement scheme where the particles
are polarized, precess in the electric field, and then their
precession is probed with high efficiency after a time on the order
of the spin-relaxation time. The noise of such one-shot measurement
is
\begin{align}
N_1 \approx \sqrt{\sN}.\label{Eq_EDM_Sens_trad_N1}
\end{align}
The corresponding S/N ratio can be improved by repeating the
measurements many times up to a total experiment time $t$:
\begin{align}
S/N \approx \frac{S_1}{N_1}\,\sqrt{\frac{t}{\tau}}= \sqrt{\sN}\,
\frac{dE}{\hbar}\,\sqrt{\tau t}. \label{Eq_EDM_Sens_trad_StoN}
\end{align}

Let us now consider the condensed-matter experiments measuring
magnetization induced by application of an electric field. Let us
say, for the sake of the argument, that we have an ideal noise-free
external magnetometer with unlimited sensitivity. What is the
statistical sensitivity of the experiment?

The EDM-induced magnetic moment of the sample is given by
\begin{align}
S_1 \propto \sN\,\frac{dE}{T} \, \mu~,\label{Eq_EDM_Sens_CM_S1}
\end{align}
where
$E$ is the effective electric field acting on the spins, $T$ is the
absolute temperature of the spins in energy units, and $\mu$ is the
magnetic moment of one spin. This is our signal. What is the noise?

In the absence of any external fields, at a given moment in time we
have a random total magnetic moment
\begin{align}
N_1 \propto \sqrt{\sN} \, \mu~.\label{Eq_EDM_Sens_CM_N1}
\end{align}
As in the case of a precession experiment, the fact that this noise
magnetic moment is random and changes in time may be used to improve
the S/N ratio. In order to characterize the correlation time of the
fluctuations, we introduce spin-relaxation time $\tau$. This
parameter characterizes how long the random magnetic moment
persists. If this time is long, this may present a serious problem
for the experiment. In other words, if the spins do not relax there
is a random signal, which would not average to near zero in a short
time.

More formally, we have expressions (\ref{Eq_EDM_Sens_CM_S1}) and
(\ref{Eq_EDM_Sens_CM_N1}) from which we can write S/N for a long
measurement time $t \gg \tau$:
\begin{align}
S/N \approx \frac{S_1}{N_1}\,\sqrt{\frac{t}{\tau}}= \sqrt{\sN}\,
\frac{dE}{T}\,\sqrt{t/\tau}~. \label{Eq_EDM_Sens_CM_StoN}
\end{align}
This shows that the key parameters for an experiment of this type
are the relaxation time $\tau$ and the temperature. Assuming that
these parameters are independent, the experiment should be done at
the lowest possible temperature to increase the degree of induced
polarization. In addition, it appears that it may be beneficial to
have \emph{fast} spin relaxation (small $\tau$), so that the
measurement can be repeated often. Such dependence of the
sensitivity on $\tau$ is the opposite of that in the case of
precession experiments [Eq. (\ref{Eq_EDM_Sens_trad_StoN})].

\section{What happens at low temperature? The usual scaling
recovered} \label{section:low temp}

Let us now consider a case where relaxation is determined by the
interaction between the spins -- the dipole-dipole interaction (see
Section \ref{subsect:Some features}). The characteristic energy
scale $\mathfrak{J}$ for such an interaction is related to the
relaxation time according to
\begin{equation}\label{Eq_EDM_Sens_J tau}
\mathfrak{J} \approx \frac{\hbar}{\tau}~.
\end{equation}
It is now important to mention that in the presence of such a
residual interaction, our assumption that the induced magnetization
is inversly proportional to the temperature breaks down when the
temperature becomes comparable to the residual interaction.
Depending on the details of the interactions, the spin system can
go, for example, into a ferro- or anti-ferromagnetic state for
$T<\mathfrak{J}$ (see, for example, Ref. \cite{Blu2003}), upon which
the susceptibility vanishes, and the system is no longer sensitive
to EDM. This effect limits the optimal temperature of the sample to
\begin{equation}
T_{opt} \approx \mathfrak{J}~.
\end{equation}
Substituting this into Eq. (\ref{Eq_EDM_Sens_CM_StoN}), and taking
into account Eq. (\ref{Eq_EDM_Sens_J tau}), we recover a result that
is identical to that of Eq. (\ref{Eq_EDM_Sens_trad_StoN}) for
``traditional" precession EDM experiments.

\section{Magnetic-field noise and the Fluctuation-Dissipation Theorem
(FDT)}

The energy associated with the spins in a polarized paramagnetic
material (this could be nuclear paramagnetism as in the case of the
Schiff-moment experiment proposed in Ref. \cite{Muk2005}) can be
written as
\begin{equation}
\sE=-{1\over 2}{\bf M}\cdot {\bf B}V,
\end{equation}
where $\bf B$ is the magnetic induction (assumed uniform in a volume
$V$) and ${\bf M}=\chi \bf B$ is the average induced magnetization;
$\chi$ is the paramagnetic susceptibility. This direct linear link
between $M$ and $B$ suggests that the fluctuations can be determined
from the FDT after we have ascertained a generalized susceptibility
(see Ref. \cite{Lan80}, Sect. 124).

The spins become polarized after application of a magnetic field due
to a dissipative process; the spins relax to the equilibrium
polarization through the \emph{longitudinal} relaxation
characterized by a time constant $T_1$.

Let us discuss the steady-state response of the magnetization to an
oscillatory magnetic field applied to the sample at a frequency
$\omega$ (with no applied static field). The specific form of the
response depends on the system. We consider two models
\begin{eqnarray}
\label{im} &&1) \ \ \ Im [\chi(\omega)]=\chi_0\frac{\omega
T_1}{1+\omega^2 T_1^2} \ ,
\\
&&2) \ \ \ Im[\chi(\omega)]=\chi_0\sqrt{\pi}\omega T_1 e^{-\omega^2
T_1^2} \ .
\end{eqnarray}
Here $\chi_0$ is the usual Curie susceptibility
\begin{equation}
\chi_0\approx \frac{\rho\mu^2}{T} \ ,\label{Eq_static_suscept}
\end{equation}
where the angular-momentum factors have been neglected, $\rho$ is
the number density, and the temperature $T$ is expressed in energy
units. The full complex susceptibility can be reconstructed using
the Kramers-Kronig relations (see, for example, Ref. \cite{Lan80},
sect. 123):
\begin{equation}
\chi(\omega)=-\frac{1}{\pi}\int_{-\infty}^{+\infty}\frac{Im[\chi(x)]
}{\omega-x+i0}dx \ .
\end{equation}
This gives
\begin{align}
&1) \ \ \ \chi(\omega) =\chi_0\frac{1}{1-i\omega T_1} \ ,\\
&2) \ \ \ \chi(\omega) =\chi_0 \left\{\begin{array}{l}
1+i\sqrt{\pi}\omega T_1-2(\omega T_1)^2+..., \ \ \omega T_1 \ll 1 \\
-1/(2\omega^2T_1^2)+..., \ \ \ \omega T_1 \gg 1 .
\end{array}
\right.
\end{align}
The first model is more relevant to electron spins and to nuclear
spins when they are coupled to the lattice (see also Refs.
\cite{Kro38,Gar59} for treatments of similar problems). The second
model is more relevant to nuclear spins in an insulator at a
sufficiently low temperature when they are decoupled from the
lattice \cite{Sli96}. We are interested in the low-frequency regime,
$\omega T_1 \ll 1$. In this regime both models give the same
results. To be specific below we use the first model \footnote{This
form of the response is analogous to the behavior of an electrical
$RC$ circuit. In the low-frequency limit $\omega T_1 \ll 1$, the
susceptibility tends to its static limit $\chi_0$, and the
magnetization is in phase with the induction. In the high-frequency
limit, $\omega T_1 \gg 1$, magnetization is $\pi/2$ out of phase
with induction, with its magnitude scaling inversely proportionally
with the frequency.}.

We are now poised to directly apply the FDT to this problem and
write an expression for the spectral density $(M^2)_\omega$  of the
square of the deviation of the magnetization from its average value:
\begin{equation} V\cdot(M^2)_\omega=\hbar
\coth(\hbar\omega/2T)\cdot Im \chi(\omega)\approx {2\chi_0 T_1
T\over 1+(\omega T_1)^2}. \label{Eq_FDT}
\end{equation}
In the last part of the above expression, we have used
$\coth(\hbar\omega/2T)\approx 2T/\hbar\omega$, which is true for $T
\gg \hbar \omega$. Combining the final expression of Eq.
(\ref{Eq_FDT}) and Eq. (\ref{Eq_static_suscept}), we get:
\begin{equation}
V\cdot(M^2)_\omega\approx \rho\mu^2 {2T_1\over 1+(\omega T_1)^2}.
\label{Eq_FDT_2}
\end{equation}
There are several properties of this expression that should be
noted. First, the average square magnetization is inversely
proportional to the volume of the sample. This represents averaging
of fluctuations over parts of a large sample. Another remarkable
result is that the magnetization noise has no temperature dependence
other than through a possible temperature dependence of $T_1$.

For a properly optimized geometry of a solid-state EDM experiment,
the detected signal depends on the magnetic moment of the entire
sample. For an experiment with an averaging time $t \gg T_1$, the
ongoing analysis reproduces the scaling of Eq.
(\ref{Eq_EDM_Sens_CM_StoN}) if we identify the relevant relaxation
time $\tau$ with $T_1$. (It is the transverse relaxation $T_2$ that
is of relevance in precession experiments.) Indeed, estimating
\begin{equation}
(M^2)_\omega\approx M^2\cdot T_1,
\end{equation}
setting $\omega=1/T_1$, multiplying both sides of Eq.
\eqref{Eq_FDT_2} by $V$, and taking the square root, we reproduce
the noise of Eq. \eqref{Eq_EDM_Sens_CM_N1}.

\section{Some features of the proposed nuclear Schiff-moment
experiment} \label{subsect:Some features}

In this section we discuss some peculiar features of nuclear
Schiff-moment experiments in ferroelectric solids proposed in Ref.
\cite{Muk2005}.

We consider a diamagnetic solid-state system with nonzero-spin
nuclei (these are the nuclear-spin $I=1/2$ $^{207}$Pb nuclei with
magnetic moment of $0.59\ \mu_N$, where $\mu_N$ is the nuclear
magneton, in the specific proposal involving ferroelectric lead
titanate). The lattice temperature is always considered cold enough,
so the effect of phonons, and specifically, the interaction between
the nuclear spins and the lattice mediated by lattice vibrations are
completely negligible. In practical terms, this would require
cooling the sample to temperatures on the order of a kelvin.

Under such conditions, the lattice is decoupled from the nuclear
spins with the exception of the fact that the spins are ``pinned" to
the lattice. Assuming that the spins only interact with each other
(by means of sensing each other's magnetic field), and that there is
no interaction with the lattice other than that the lattice keeps
the nuclei fixed in space, it is straightforward to estimate the
spin-relaxation rate (see, for example, Ref. \cite{Kit2005}, Ch.
13). Because magnetic field from a dipole falls as the inverse third
power of the distance, for a given spin, relaxation is determined by
its closest neighbor(s). The relaxation rate can be estimated as the
Larmor precession rate of a spin in its neighbor's field:
\begin{align}
\gamma \sim \frac{(\mu_N)^2}{\hbar r^3}~.
\end{align}
Here $\mu_N$ is the nuclear magneton, and $r$ is the characteristic
distance between the neighbors. If the distance between interacting
spins is on the order of interatomic spacing in condensed matter,
the relaxation rate is on the order of kilohertz. This relaxation
provides a lower limit on the magnetic-resonance linewidth. For the
specific case of lead titanate, the dipole-dipole relaxation rate is
estimated in Ref. \cite{Muk2005} as being $\gamma/(2\pi)\approx 200\
$Hz. It is important that despite the fact that the nuclear spins
are isolated from the lattice, the total angular momentum of the
nuclear spin-system is not conserved. This is easy to see from the
following argument involving, for simplicity, just two spins.

The Hamiltonian describing the interaction between the spins is
\begin{align}
\hat{H}=-\vec{\mu}_1 \cdot \vec{B}_{21} =-\vec{\mu}_1 \cdot
\frac{3\prn{\vec{\mu}_2\hat{r}_{12}}\hat{r}_{12}-\vec{\mu_2}}{r_{12}^3}~.\label{Eq:DD_relaxation_Ham}
\end{align}
Here $\vec{\mu}_{1,2}=g_{1,2}\mu_N\vec{I}_{1,2}$ are the magnetic
moments of the two spins, $g_{1,2}$ are their nuclear $g-$factors,
$\vec{I}_{1,2}$ are their spin operators, $\vec{r}_{12}$ is the
separation between the spins, and $\hat{r}_{12}$ is the unit vector
in the direction of $\vec{r}_{12}$.

Let us examine whether the total spin projection $M_1 + M_2$ onto a
given quantization axis is a conserved quantity. To do this, we
check whether the corresponding operator $I_z=I_{1z}+I_{2z}$
commutes with the Hamiltonian of Eq.(\ref{Eq:DD_relaxation_Ham}).
\begin{align}
&\sbrk{I_z,\hat{H}}=\label{Eq:DD_relaxation_Comm}\\
&\frac{-g_1g_2\mu_N^2}{r_{12}^3}\sbrk{I_{1z}+I_{2z},3\prn{\vec{I_1}\cdot\hat{r}_{12}}\prn{\vec{I_2}\cdot\hat{r}_{12}}-
\vec{I}_1\cdot\vec{I}_2 }~.\nonumber
\end{align}
\medskip

The commutator term $\sbrk{I_{1z}+I_{2z},\vec{I}_1\cdot\vec{I}_2 }$
is zero, but the other term in Eq.(\ref{Eq:DD_relaxation_Comm}) is
generally not. This is because, for example
$\vec{I_1}\cdot\hat{r}_{12}$ is a linear combination of the
operators $I_{1x}$, $I_{1y}$, and $I_{1z}$ the first two of which do
not commute with $I_{1z}$.

Thus we see that the total spin angular momentum is not conserved in
dipole-dipole interactions, and the angular momentum is exchanged
with the lattice. A detailed discussion of the evolution of systems
of many spins on a lattice has been given in Ref. \cite{Sod95}.

The scale of the dipole-dipole interaction strength $\mathfrak{J}$
expressed in temperature units corresponds to tens to hundreds of
nanokelvin. As discussed in Section \ref{section:low temp}, the EDM
experiment would ideally be conducted at spin temperatures slightly
higher than this.

At this point, prior to proceeding with the discussion of the EDM
measurement, let us consider several thought experiments that will
help in understanding of the spin system.

We first assume that a strong magnetic field is initially applied,
so the spins are polarized. (It is not necessary that the field be
strong enough to lead to full polarization, but it has to be much
stronger than the characteristic value of the dipole field.) We then
turn off the leading field abruptly. The question is: to which state
does the system relax, and at what rate?

The way we have set up the problem, the magnetic interaction between
the spins is the only interaction affecting the spins, so the spin
polarization will relax at a rate on the order of $\mathfrak{J}$
(where we do not distinguish between energy, temperature, and
frequency units). Because, as discussed above, angular momentum is
not conserved, the final state of the spins will have no average
polarization. The temperature of the spins will remain the same. In
this state, each of the spins ``sees" a randomly fluctuating field
from other spins, which has a characteristic correlation time of
$1/\mathfrak{J}$ and just the appropriate characteristic magnitude
that it rotates the spin (via Larmor precession) by an angle of
order unity during a correlation time. Consequently, the overall
magnetic moment randomly oscillates with the same correlation time,
and the overall magnitude proportional to the square root of the
total number of spins in the sample as discussed in the preceding
sections. These time-dependent fluctuations are essential for the
Schiff-moment experiment as they serve to average the random
polarization of the sample, while preserving the ``bias" due to the
P,T-odd effect.

An interesting question is what happens if the strong magnetic field
is reapplied quickly (much faster that the correlation time)? After
the field is turned on, the magnitude of this strong field is much
greater than the dipole fields, and each of the spins precesses
around the direction of the strong field. Effectively, in this
regime, the components of the dipole fields perpendicular to the
leading field have no effect on the spins, and the only effect of
the longitudinal components is to produce a small variation of the
overall field magnitude from site to site. Such inhomogeneous
broadening is important for transverse ($T_2$) relaxation, but is
irrelevant for longitudinal ($T_1$) relaxation. Thus, after the
application of the strong field, the spin system remains in the
unpolarized state \footnote{More precisely, the system remains in a
state with a random weak polarization along the leading field equal
to the polarization component in this direction that existed due to
fluctuations at the moment when the strong field was turned on.}
indefinitely, in the framework of the approximations that we have
assumed here. In practice, some slow $T_1$-relaxation processes will
eventually relax the spins into a state where their magnetic moments
are preferentially along the strong leading field, which is the
equilibrium state.  Note that such behavior of the nuclear-spin
subsystem isolated from the lattice has been discussed already half
a century ago in Ref. \cite{Heb59}.

Next, we discuss how the nuclear spin-system can be cooled to a low
temperature (the desired temperature is slightly above
$\mathfrak{J}$). This will require slow decrease of a leading field
as opposed to rapid leading-field variations. Suppose an
experimentally realizable magnetic field of $B=10^5\ $G is applied,
and the sample is cooled down to a temperature $T_0 \sim 1\ $K where
the nuclear spins decouple from the lattice. At this point, the
polarization for the case of $^{207}$Pb is
\begin{align}
\approx\frac{\mu B}{T_0}\sim 10^{-3}.\label{eq:small_factor}
\end{align}
The magnetic field is then slowly turned off causing
adiabatic-demagnetization cooling of the nuclear spin-system. The
spin temperature at the end of cooling can be estimated as
\begin{align}
T\sim T_0 \frac{J}{\mu B}=T_0\cdot\prn{\frac{\mu B}{T_0}}^{-1}\cdot
\frac{\mathfrak{J}}{T_0} \sim 10^{-5} -
10^{-4}\textrm{K}.\label{Eq_Spin_T_est}
\end{align}
Unfortunately, due the smallness of the factor
(\ref{eq:small_factor}), this is a significantly higher temperature
than the desired $\gtrsim \mathfrak{J}$.

\section{Estimate of the statistical sensitivity of the Schiff-moment
experiment}

Let us take the nuclear spin temperature $T=10^{-4}\ $K, a
conservative estimate in Eq. (\ref{Eq_Spin_T_est}). The magnetic
moment of a ferroelectric lead-titanate sample induced by a Schiff
moment, according to Eq. (8) of Ref. \cite{Muk2005}, is
\begin{align}
V\cdot M \approx 10^6\ \sN \mu S\ \frac{1\ \textrm{eV}}{T}\sim
10^{14}\ \sN \mu S.
\end{align}
Here the Schiff moment $S$ of the $^{207}$Pb nucleus should be
expressed in units of $e\cdot a_0^3$.

Estimating the signal-to-noise ratio (assuming noise-free
magnetometer) along the lines of the discussion in Section
\ref{Section_High-Temp_Limit}, we have
\begin{align}
S/N\sim 10^{14}\sqrt{\sN}S\sqrt{\gamma t}\sim
10^{30}S.\label{Eq_Sch_Est_rev}
\end{align}
For the final step of the estimate (\ref{Eq_Sch_Est_rev}), we have
taken $\sN=3.3\cdot 10^{22}$ corresponding to a volume of $V=10\
$cm$^3$ and the natural abundance of $^{207}$Pb; the experiment
duration of $t=10\ $days, and $\gamma\approx 10^{-12}$eV (in
frequency units, $\gamma/(2\pi)\approx 200\ $Hz). Thus, an $S/N=1$
corresponds to a sensitivity to the Schiff moment of approximately
$10^{-30}\ e\cdot a_0^3$. This is by more than four orders of
magnitude better than the present best limits on the Schiff moment
of  $^{199}$Hg (see Ref. \cite{Rom2001} and references therein).

Finally, it is interesting to estimate a characteristic magnitude of
the spin-noise magnetic field. Assuming a sample with all
characteristic dimensions $2R$, just outside of it, the noise
magnetic field is on the order of
\begin{align}
B_N \sim \frac{\sqrt{\sN}\mu}{R^3}\sim 10^{-12}\
\rm{G}.\label{Eq_B_N}
\end{align}
The noise produced by the spins is comparable to the noise of modern
magnetometers, see, for example, Ref. \cite{Kom2003} and references
therein.

\section{Conclusion}

In this note, we have considered the EDM experiments that rely on
measuring magnetization of a condensed-matter sample induced by
application of an electric field. A limit on statistical sensitivity
of such an experiment arises due to random fluctuations of the spins
in the sample. We find that, while the ultimate sensitivity has the
usual scaling ($\propto \sqrt{\sN t}$) with the number of spins and
the measurement time, in the limit where the temperature greatly
exceeds the spin-spin interaction energy, the sensitivity also
scales $\propto \sqrt{\tau}/T$. Such scaling with relaxation time is
radically different from that for the more traditional precession
EDM experiments. Interestingly, the usual scaling is recovered if
one is able to cool the spins to a low temperature, comparable to
the energy of the magnetic dipole-dipole interaction between the
spins.

After presenting a heuristic derivation of this result, we have
discussed how it can be obtained from the fluctuation-dissipation
theorem.

Finally, we have presented an estimate based on the earlier results
of Ref. \cite{Muk2005} combined with the present considerations of
the noise due to spin fluctuations of the ultimate statistical
sensitivity of a search for the P- and T-odd nuclear Schiff moment
using a ferroelectric material. We find that, with realistic
experimental parameters, the statistical noise due to spin
fluctuations should not preclude obtaining a significant improvement
in sensitivity to the Schiff moment (perhaps, up to four orders of
magnitude) compared with the present best limits.

An important limiting factor for the nuclear Schiff-moment
experiment appears to be the difficulty of cooling the spin system
to a sufficiently low temperature (in the tens of nanokelvin range)
using the adiabatic demagnetization technique. The limitation comes
from the fact that thermal polarization of the spins in an
achievable laboratory magnetic field is very low. In principle, it
may be possible to produce much higher initial nuclear-spin
polarizations, for example, by creating UV light induced metastable
paramagnetic centers \cite{War96} and performing optical pumping.

\begin{acknowledgments}
We are grateful to E.D. Commins, B. V. Fine, J. Haase, E.L. Hahn, M.
G. Kozlov, M. Ledbetter, J. E. Moore, M. V. Romalis, C. P. Slichter,
and M. Zolotorev for helpful discussions. This work was supported in
part by the UC Berkeley-LANL CLC program and by NSF.
\end{acknowledgments}

%

\end{document}